\documentclass[journal]{vgtc}                
\ifpdf
  \pdfoutput=1\relax                   
  \usepackage{graphicx}                
  \DeclareGraphicsExtensions{.pdf,.png,.jpg,.jpeg} 
\else
  \usepackage{graphicx}                
  \DeclareGraphicsExtensions{.eps}     
\fi%

\graphicspath{{figures/}{pictures/}{images/}{./}} 

\usepackage{microtype}                 
\PassOptionsToPackage{warn}{textcomp}  
\usepackage{textcomp}                  
\usepackage{mathptmx}                  
\usepackage{times}                     
\usepackage{cite}                      
\usepackage{tabu}                      
\usepackage{booktabs}                  

\usepackage{caption}

\usepackage{dblfloatfix}    

\onlineid{1238}

\vgtccategory{Research}
\vgtcpapertype{application/design study; evaluation}

\title{Investigating Search Among Physical and Virtual Objects Under Different Lighting Conditions
}

\author{You-Jin Kim*, Radha Kumaran*, Ehsan Sayyad, Anne Milner, Tom Bullock, Barry Giesbrecht,\\and Tobias Höllerer, Senior Member, IEEE}
\authorfooter{
\item
 You-Jin Kim, Ehsan Sayyad and Tobias Höllerer are with Media Arts and Technology, University of California, Santa Barbara.
\item
 Radha Kumaran and Tobias Höllerer are with the Department of Computer Science, University of California, Santa Barbara.
 \item Anne Milner, Tom Bullock and Barry Giesbrecht are with the Department of Psychological and Brain Sciences and the Institute for Collaborative Biotechnologies, University of California, Santa Barbara.
 \item Kim and Kumaran contributed equally to this paper.

 \item
This is a preprint version of this article. The final version of this paper can be found in the IEEE TVCG (Vol. 28, 2022). For citation, please refer to the published version. This work was initially made available on the author's personal website [yujnkm.com] in October 2022, and was subsequently uploaded to arXiv for broader accessibility.
}

\shortauthortitle{Kim \MakeLowercase{\textit{et al.}}: Search Among Physical and Virtual Objects Under Different Lighting Conditions}

\abstract{By situating computer-generated content in the physical world, mobile augmented reality (AR) can support many tasks that involve effective search and inspection of physical environments.  Currently, there is limited information regarding the viability of using AR in realistic wide-area 
outdoor environments and how AR experiences affect human behavior in these environments. Here, we conducted a wide-area outdoor AR user study ($n$=48) using a commercially available AR headset (Microsoft Hololens 2) to compare (1) user interactions with physical and virtual objects in the environment (2) the effects of different lighting conditions on user behavior and AR experience and (3) the impact of varying cognitive load on AR task performance.  Participants engaged in a treasure hunt task where they searched for and classified virtual target items (green ``gems") in an augmented outdoor courtyard scene populated with physical and virtual objects. Cognitive load was manipulated so that in half the search trials users were required to monitor an audio stream and respond to specific target sounds.  Walking paths, head orientation and eye gaze information were measured, and users were queried about their memory of encountered objects and provided feedback on the experience. Key findings included (1) Participants self-reported significantly lower comfort in the ambient natural light condition, with virtual objects more visible and participants more likely to walk into physical objects at night; (2) recall for physical objects was worse than for virtual objects, (3) participants discovered more gems hidden behind virtual objects than physical objects, implying higher attention on virtual objects and (4) dual-tasking modified search behavior.  These results suggest there are important technical, perceptual and cognitive factors that must be considered if the full potential of ``anywhere and anytime mobile AR" is to be realized.} 


\teaser{
  \centering
  \includegraphics[width=1.0\textwidth]{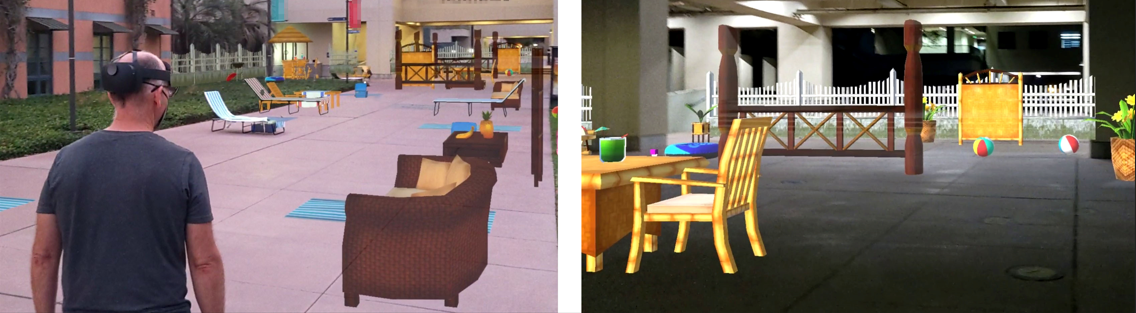}
  \caption{Left: Participant navigating the augmented outdoor environment in the natural lighting condition. Right: Environment during nighttime condition. HoloLens-2 MR capture does not depict relative brightness and opacity true to the actual experience.}
  \label{fig:teaser}
}

\renewcommand{\manuscriptnotetxt}{}

\vgtcinsertpkg

\begin{document}


\firstsection{Introduction}

\maketitle

Augmented reality (AR) is widely seen as an ingredient, if not a central paradigm, in next-generation mass-adopted mobile information consumption. Most major technology companies have demonstrated steps or have paid lip service towards AR as a platform for complementing or replacing mobile hand-held devices as the main provider of context-based information. With head-worn AR displays, which are still somewhat cumbersome to wear today but are ultimately projected to shrink to the size of standard prescription eyewear, users would have access to meaningful world-registered context information at any time, wherever they go. This vision of anywhere and anytime head-worn AR is compelling, but the hoped-for generality of use is not a given, or easily facilitated by state-of-the-art system design. Impressive progress has been made over the past two decades on wide-area simultaneous localization and mapping (SLAM), and AR-cloud-based aggregation of AR tracking environments via shared ``anchors" is now supported by several competing services (Google, Microsoft, Apple, Facebook). However, tracking performance and robustness in wide-area unconstrained environments, and especially across different lighting conditions, is still a considerable challenge. Even if tracking were fully solved, it is unknown how ambulatory users will react to AR stimuli in different situations, while walking in large outdoor spaces, and under different lighting conditions. 

There is limited information on the impact of constant AR availability and use in realistic wide-area outdoor settings on participant behavior in scene exploration scenarios. This paper presents the first user study on AR search behavior in a large augmented outdoor environment. Today's commercial AR headsets afford wide-area tracking in carefully set up environments, and the goal of this study was to utilize the full extent of tracking and display capabilities of state-of-the-art commercial headsets to explore the effects of head-worn AR experiences on participant search behavior and subjective user experience. 

\begin{figure*}[thb]
 \includegraphics[width=0.24\textwidth]{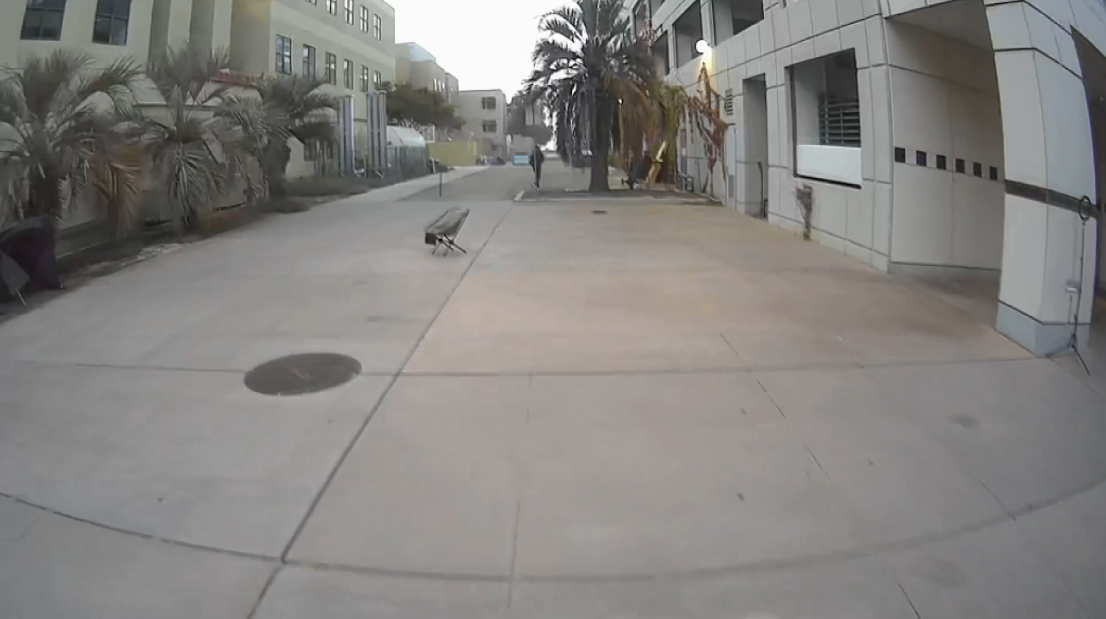}\hfill
 \includegraphics[width=0.24\textwidth]{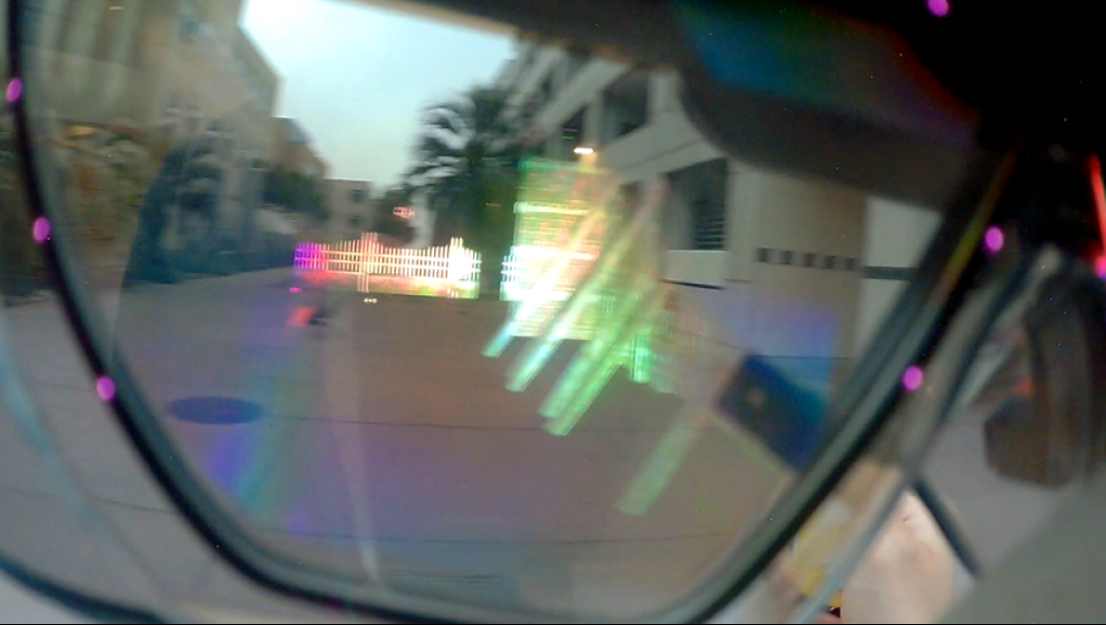}\hfill
 \includegraphics[width=0.24\textwidth]{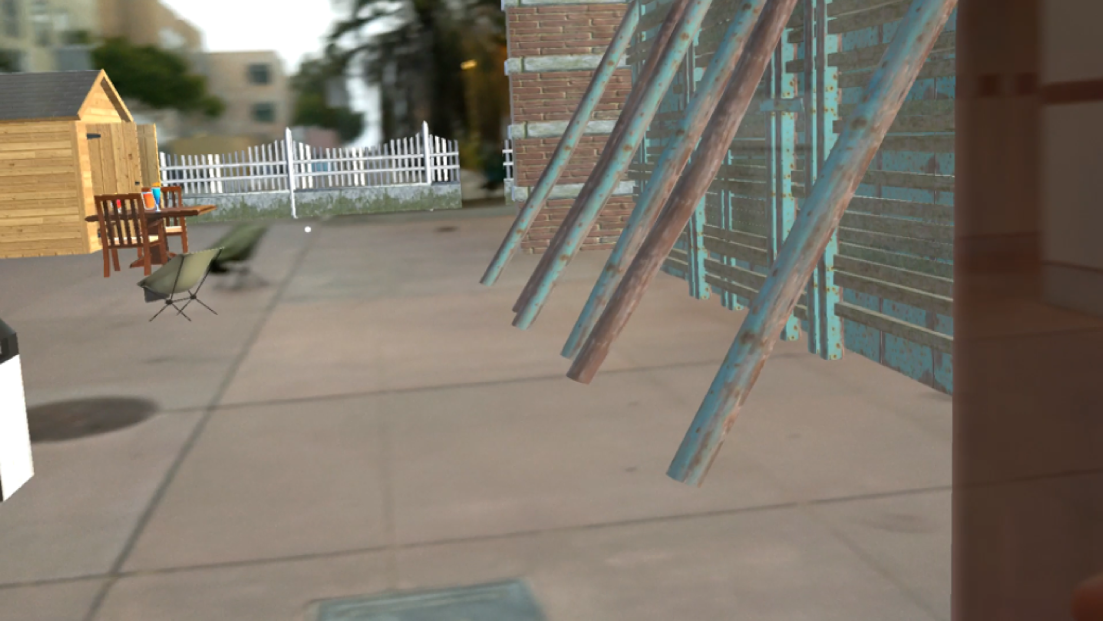}\hfill
 \includegraphics[width=0.23\textwidth]{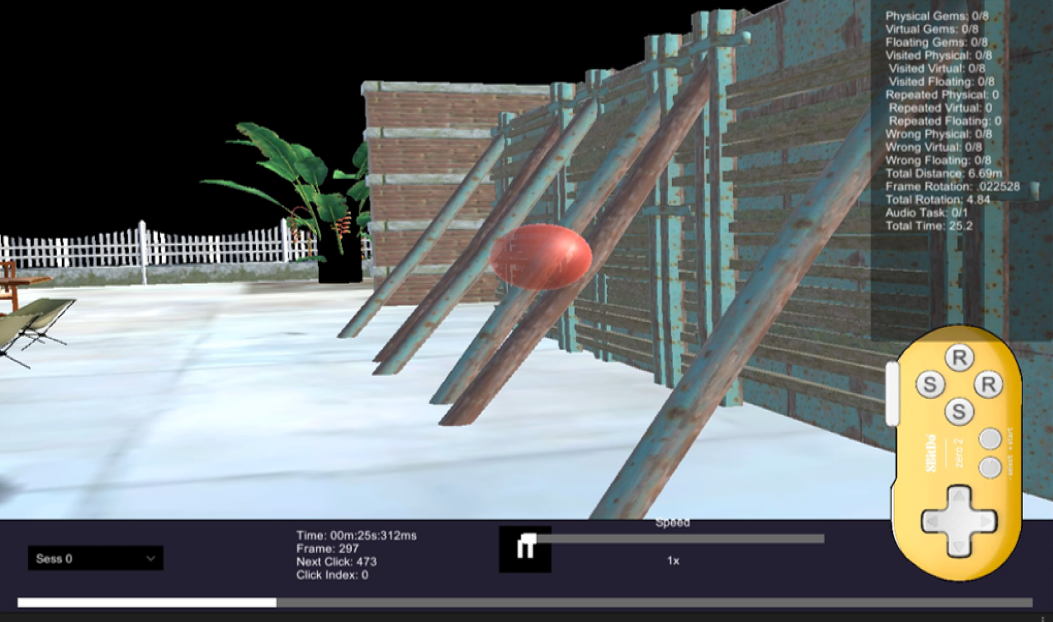}\hfill
 \caption{Captured user experience. From left to right: 1) Video frame from added external camera. 2) View through HoloLens-2. Eye tracking LEDs are not perceivable to participant, and view more resembles 3) HoloLens-2 MR Capture (but has a slightly smaller augmented field of view). 4) All our captured information can be used to reconstruct and play back or scrub through the entire user experience in our playback tool.}  
 \label{fig:dataCollection}
\end{figure*}

Our first goal was to assess the influence of physical and virtual objects on AR users' behavior in a wide-area search and classification task. AR users will have to adapt to a blend of physical and virtual objects in their visual environment, so it is important that differences in how they interact with both types of objects are well characterized and understood.  Our second goal was to test the impact of different lighting conditions on users' behavior and subjective reports of the AR experience.  Mass adoption of AR will presumably entail its use under a range of different lighting conditions, and during our initial pilot testing with the HoloLens-2 headsets it became apparent that a) the lighting situation for our outdoor studies had to be constrained to lighting conditions that did not consist of direct sunlight, and b) that user perception of the augmented world likely would differ considerably between ambient natural light and nighttime settings. Accordingly, lighting condition was included as an additional independent variable in the design.  Our third goal was to assess the impact of cognitive load on users' performance.  AR systems will be used by individuals while under varying states of cognitive load (e.g. navigating the visual world using AR cues while also talking on the phone) so it is important to understand how multi-tasking impacts attention to the AR environment.

The setting for the present study was an augmented outdoor environment populated with physical and virtual objects evoking a leisure-oriented courtyard scene, such as lawn and lounge chairs, umbrellas, picnic tables, and assorted beach paraphernalia (cf. Figures \ref{fig:teaser} and \ref{fig:dataCollection}). Participants were asked to complete a set of search tasks that involved finding and classifying virtual gems that were distributed, and to some extent hidden, behind the physical and virtual objects in this environment. To study the effects of different lighting conditions we ran participants during ambient natural light (during the evening, to avoid direct sunlight and maintain consistent tracking and display operation) and at nighttime under artificial light (scene lit by lamplight).  To manipulate cognitive demands, we required that all participants complete an additional auditory stimulus response task on half of the trials, so that half the experiment was completed under single-task and half under dual-task conditions.  


We predicted that performance would be degraded in the virtual gem search and classification task when participants were also required to perform the secondary audio stimulus task concurrently.  It was not possible to make any strong predictions regarding the effects of lighting conditions on gem search and classification, but based on previous AR experiences with different brightness modes~\cite{erickson2020review}, we surmised that gem search might be more successful at night because the gems are more visually salient relative to the surrounding environment.  It was also not possible to make any predictions regarding the impact of different lighting conditions on various aspects of the AR user experience e.g. satisfaction, comfort, visual fatigue.



To the best of our knowledge, this is the first outdoor AR user study examining the effects of lighting situation on user experience, and at the same time also the largest-area controlled headworn-AR user study outdoors.  Some of our key results and novel insights include:

\begin{itemize} 
    \item Participant responses indicated they found the tasks more demanding in the natural light condition, with significantly lower comfort ratings and a trend towards more fatigue. Despite efforts to match absolute display brightness across the lighting conditions, participants rated virtual objects more visible at night, real objects more visible during the day, and reported a greater likelihood of walking into physical objects at night.
    
    \item Participants recalled physical objects in the environment less than virtual objects, suggesting a stronger awareness of the virtual environment than the physical one.
    
    \item Participants were more successful at finding and classifying free-floating gems than gems hidden behind virtual or physical objects, and they were more accurate at detecting gems behind virtual objects than physical objects.
    
    \item Participants’ total head rotation was reduced during the dual-task condition, suggesting that the increased cognitive load impacted search behavior.
\end{itemize}

\section{Related Work}\label{relatedwork}
Here we discuss related prior work in four categories: wide-area AR, perceptual effects in AR, cognitive map building, and dual-task cognitive load experiments.

\subsection{Wide-Area AR}
From the earliest days of Augmented Reality, using AR in wide areas has been a main focus in the field\cite{thomas1998a,newman2001augmented,hollerer2004mobile}. Since AR faces many technical challenges such as tracking, rendering and optics, up until recent years performing augmented reality user studies in a comprehensive way would require strictly controlled environments and heavy equipment\cite{vlahakis2001archeoguide, gleue2001design}. Even with these challenges, there were some groundbreaking early works exploring wide-area AR\cite{thomas2002first,cheok2004human,gabbard2007active}.  Due to these limitations, most of the related work in Wide-Area AR falls under either hand-held mobile AR or Virtual Reality\cite{morrison2009like, sayyad2020walking}. Vroamer \cite{cheng2019vroamer} and DreamWalker \cite{yang2019dreamwalker} are dynamic virtual reality systems that use real-world data and replace that with a complete virtual environment, letting the user to navigate a wide-area space. They both include task completion and post study questionnaires to validate the design choices. With advances in visual inertial odometry and the introduction of affordable mobile augmented reality HMDs such as Microsoft HoloLens and Magic Leap, there are available opportunities to explore user behavior in larger areas, under longer time limitations and with more versatile lighting conditions. Libraries like Vuforia, Google's ARCore and Apple's ARKit, have made developing handheld mobile augmented reality applications easier and less costly. Many companies such as Niantic started their proprietary SLAM software to use this technology across their products. Games like Pok\'{e}mon Go are a product of these efforts. Many augmented reality user studies rely on these widely played games\cite{Hamari2019UsesAG,Alha2019WhyDP}. Other work has explored large-scale AR environments for collaborative experiences \cite{rompapas2019towards, rompapas2018holoroyale}.

\subsection{Perceptual Effects in Augmented Reality}
Viewing a complex physical environment while concurrently viewing a superimposed virtual scene can result in a high perceptual and cognitive load. Currently the visual perceptual effects while using an AR headset are not fully understood. Prior research has demonstrated that distance judgements of physical objects are typically underestimated while viewing an AR scene \cite{swan2007egocentric, smith2015optical}.  Previous work studying text readability\cite{gattullo2014effect} and color schemes\cite{kim2019effects} in an indoor AR setting suggest that the physical environment lighting could affect user performance and preferences. Color can also play an important role in both perceiving a virtual object correctly, perceiving depth-cues \cite{gabbard2010more} and visual acuity\cite{livingston2013basic} while the lack of focus cues in head-mounted displays could affect perception of virtual objects \cite{condino2019perceptual}. Furthermore, under changing light conditions the chosen colors of virtual objects may cause perceptual problems\cite{newman2001perceptual, kahl2022influence}. Our work looks at the effect of physical lighting in an outdoor AR environment on user experience and performance.

\subsection{Cognitive Map Building}

When first introduced to a new environment, humans quickly learn information pertaining to the spatial layout of the environment such as locations, distances, and directions \cite{montello1998new}. Rapid learning of a new environment is achievable because of the ability to form a cognitive map from the outset. A cognitive map is a mental representation of the spatial environment to guide navigation and support memory for places previously visited \cite{ruddle2011walking}. Individual differences in spatial ability rely on working memory as spatial layouts require maintaining information from multiple views \cite{montello1993scale}. Self-reported sense of direction has been shown to be a simple measure of spatial ability which is determined by accuracy of an individual’s mental representation \cite{kozlowski1977sense}. Furthermore, self-reported sense of direction has previously been shown to reliably predict way-finding performance during navigation \cite{hund2009effects, hegarty2006spatial}. The encoding of spatial information is another important factor in forming high-quality cognitive maps. Vision is the primary sensory modality that humans utilize to encode spatial information. However, proprioception, the movement of the body through an environment, plays an important role independently from vision \cite{waller2004body}. Proprioception has been found to significantly improve navigational performance and accuracy of cognitive maps \cite{ruddle2009benefits, ruddle2011walking}. Therefore, walking within an environment should result in better cognitive map building compared to non-translational locomotion techniques. In our work we explore the effects of proprioception and sense of direction on user behavior and performance in a wide-area search and classification task.

\subsection{Dual-Task Cognitive Assessment}
Performing multiple tasks simultaneously often occurs in everyday life. The dual-task paradigm involves concurrently engaging in two tasks and examining the interference of one or both tasks on one another. Due to limited attentional resources, performing multiple tasks can lead to an impairment in performance which is typically observed as slower reaction times (RTs) and reduced accuracy \cite{pashler1989chronometric}. It has been argued that these limitations can be reduced if the two tasks involve different sensory modalities such as vision and audition as these rely on independent attentional resources \cite{wickens1984processing}. However, it has been demonstrated that performance in a visual task can be reduced when simultaneously attending to an auditory task suggesting that performance can indeed be impaired to some extent when the dual-tasks are involve two different modalities \cite{spence1997audiovisual}. The negative impact of cell phone use while driving has been attributed to attentional resources being directed towards an auditory stream which distracts selective visual attention leading to poor driving performance \cite{strayer2001driven}. These studies demonstrate that attending to an auditory stream can have a negative impact on visual performance. Additionally, AR headset's binocular conditions impact the accuracy of manual tasks\cite{condino2019perceptual}. In the current study, we examined performance under single-task and dual-task conditions where the two tasks consisted of a visual search task and an auditory task both delivered via an AR headset. 

\begin{figure}[t]
 \centering \includegraphics[width=0.99\columnwidth]{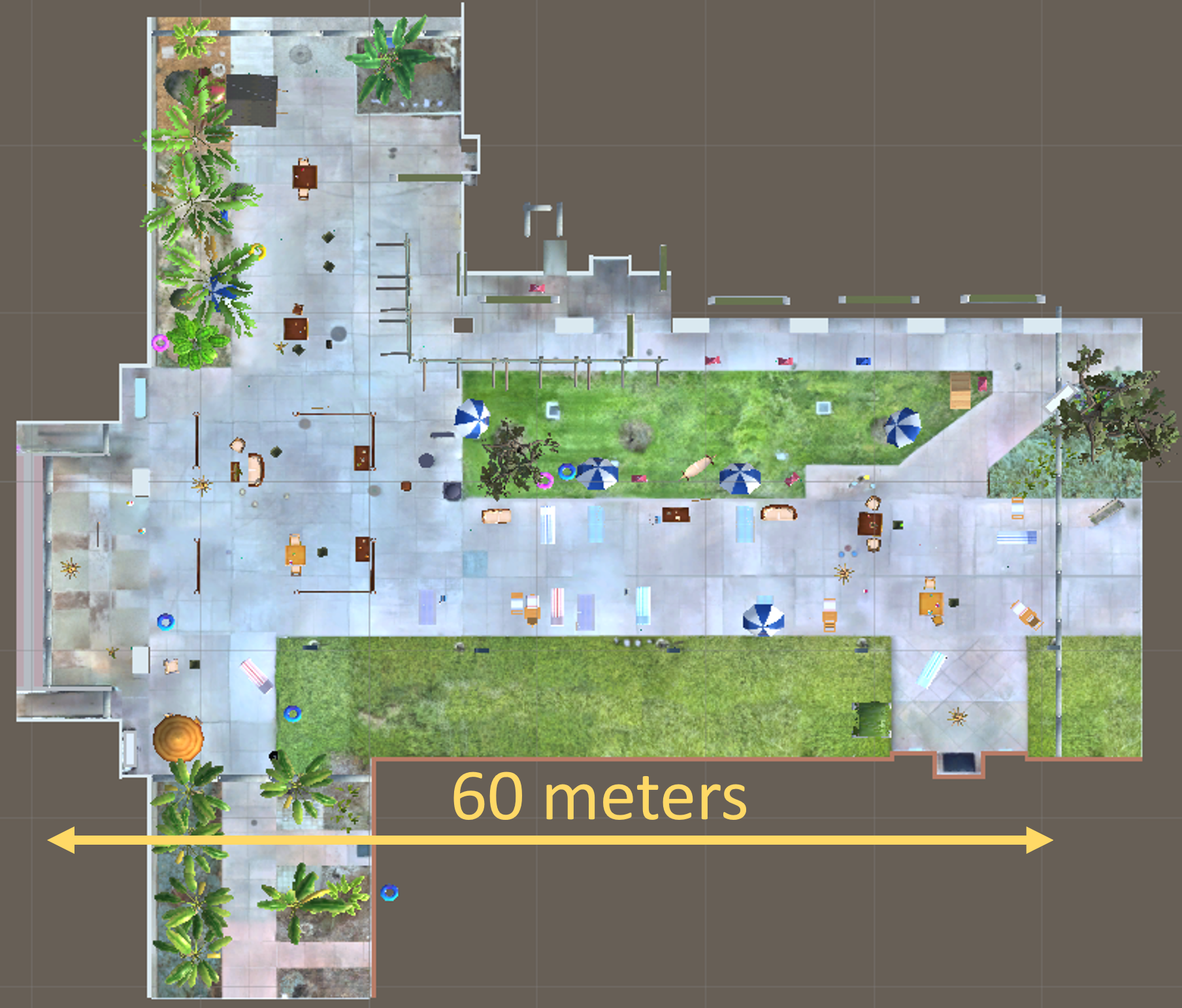}

 \caption{Our study environment is completely modeled in 3D, which is used to handle occlusions on device, design the experiment layout, and replay the captured user data to re-assess each trial offline. This top-down map shows the extent of our walkable areas and the physical and virtual augmentations for one particular trial.}
 \label{fig:map}
\end{figure}

\section{Wide Area Outdoor AR: Environment and Formative Experimentation}

\label{sec:formativeExp}

    \paragraph{Direct Sunlight} When considering the duration of the experiment as well as the most optimal time to conduct it, we observed that it was significantly harder to see the virtual environment in directly lit areas with bright sunlight. We experimented with tinted foil to cover the visor~\cite{lages2019enhanced}. This method highlighted the virtual objects, but altered the overall visibility and experience. Therefore we conducted the experiments at dusk, when there was enough natural ambient light but no direct sunlight, to avoid decline of usability caused by illumination intensity~\cite{kahl2022influence}. This condition is referred to as ``natural light".  We also observed an improvement in tracking performance at this time of day, which further solidified our choice.

    \paragraph{Tracking Performance}
    Since HoloLens is optimized for indoor environments\cite{microsoftmapphysical}, persistent tracking and spatial mapping in a large outdoor area is challenging. We used Microsoft's “World Locking Tools” (WLT). 
    Since we couldn't use Hololens's spatial mapping to handle occlusions effectively in our large outdoor setup, we modelled a virtual replica of our physical space to handle occlusion. We used 6 WLT Space Pins to align our virtual replica of the space with the physical space. This enabled us to design the experiment completely in desktop. We used an SFM 3D reconstruction of our experiment environment to model the virtual replica of the area and correctly place the virtual objects in relation to the physical space (see Figure \ref{fig:map}).

\begin{figure}[t]
 \centering \includegraphics[height=4.35cm]{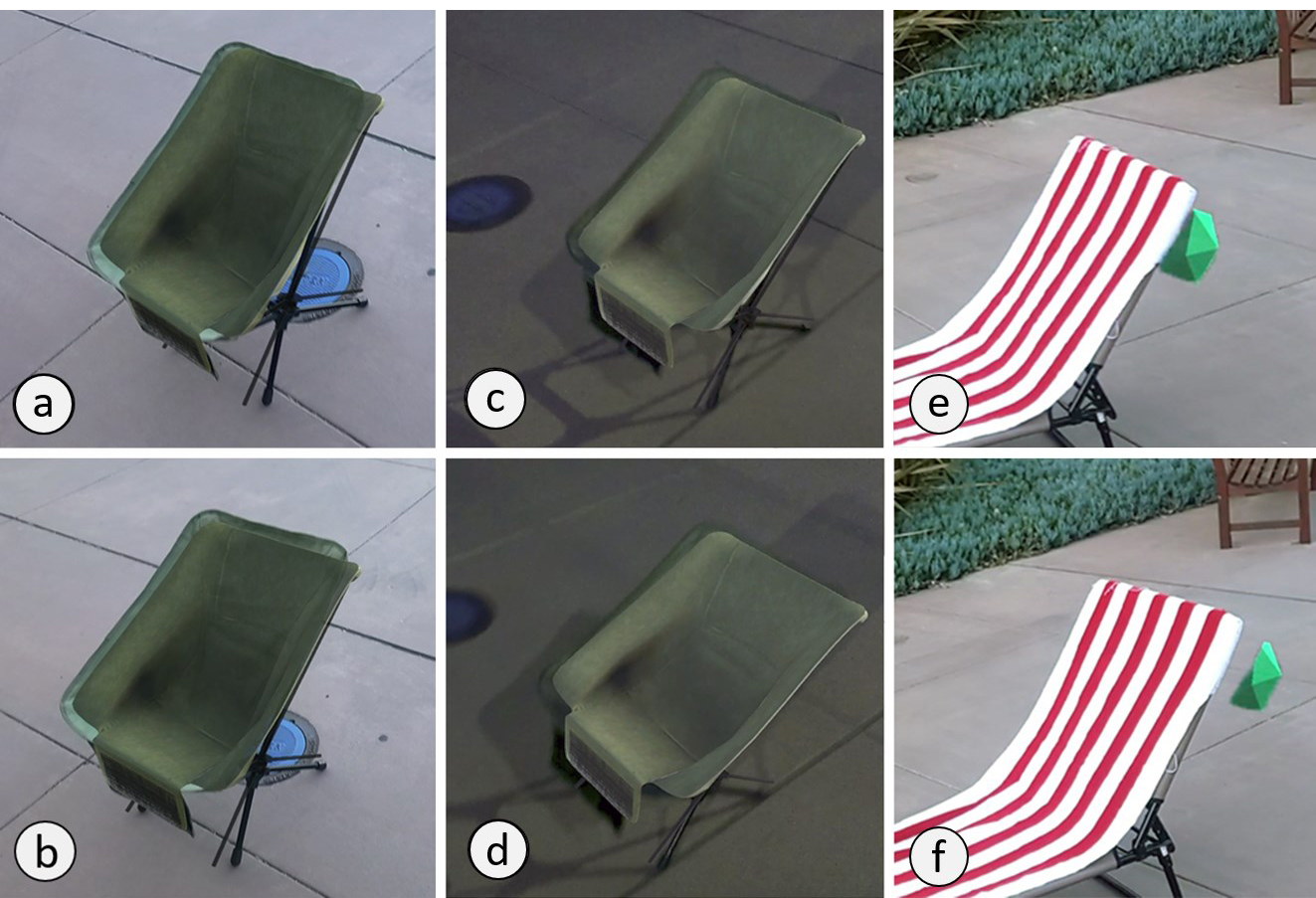}
 \caption{View of virtual objects superimposed upon real counterparts (in the expected position), and gems occluded by real objects. (a) Natural light, as soon as the environment is aligned. (b) Natural light, after walking around the environment for five minutes. (c) Night time, as soon as the environment is aligned. (d) Night time, after walking around the environment for five minutes. (e) Gem occlusion with perfect tracking. (f) Gem occlusion with drift.} 
 \label{fig:trackingChange}
\end{figure}

    One set of WLT Space Pins were uploaded to the cloud for each lighting condition. Upon launch, the software loaded the respective space pins from the cloud and the virtual scene was aligned to the real world without intervention. Stable outdoor internet was required, which was necessary to create, store and retrieve the anchors used in the world-locking system. We cancelled studies on days when stable internet was not available. This happened twice. Occasionally, some pins would require re-adjustment after loading. Experimenters aligned the environment at the beginning of the study and it usually stayed stable for the duration of the study. Although objects would sometimes appear to be misaligned when viewed from a distance, moving closer to the objects improved the alignment, as seen in Figure \ref{fig:trackingChange}. Participants were instructed to notify the experimenters if they experienced visual registration problems, and we also identified such occurrences using our playback software after the fact (we reviewed every single trial at least twice in playback). The small time intervals that were affected by temporary tracking disengagement were excluded from gem classification and behavioral metrics analysis. There were times when tracking disengaged to the point that the experimenter had to intervene and re-align the augmentations. Any trial during which this happened, was discarded and a backup trial was run at the end of that user’s study time. 39 out of the 288 total trials (6 (trials) x 48 (participants)) were such repeated trials. Participants were asked about the stability of the virtual environment in the post-questionnaire and 60.3\% of them responded with a score $\geq$5 (on a scale of 1-7, with 7 being very stable). Only 20.7\% of participants answered with a score of 3 or less, with no participants rating below 2.


    
    
\section{Experiment}

In this experiment, we analyze the effects of environmental lighting, target location (among virtual and physical objects), and a dual-task condition on outdoor AR  search and classification task performance.

\subsection{Design}

Three independent variables were manipulated: task, gem location and lighting.

    \paragraph{Task} There were two task conditions. In one condition, participants were required to find and classify gems at different locations in the environment (see Gem location). In the second condition, participants were required to perform the gem classification task, as in the first condition, but in addition, they were required to do an audio task. The audio response task required participants to listen to a randomized sequence of five words from the NATO phonetic alphabet (e.g., Alpha, Bravo, Charlie, Delta, Echo) and respond to each occurrence of a predefined target word unique to each participant by clicking a specified button on the controller. The stream of words was presented continuously and there was a 2-5 second delay between subsequent words. The two audio task conditions are referred to as ``audio absent" and ``audio present", respectively.

\begin{figure}[t]
 \centering \includegraphics[height=4.35cm]{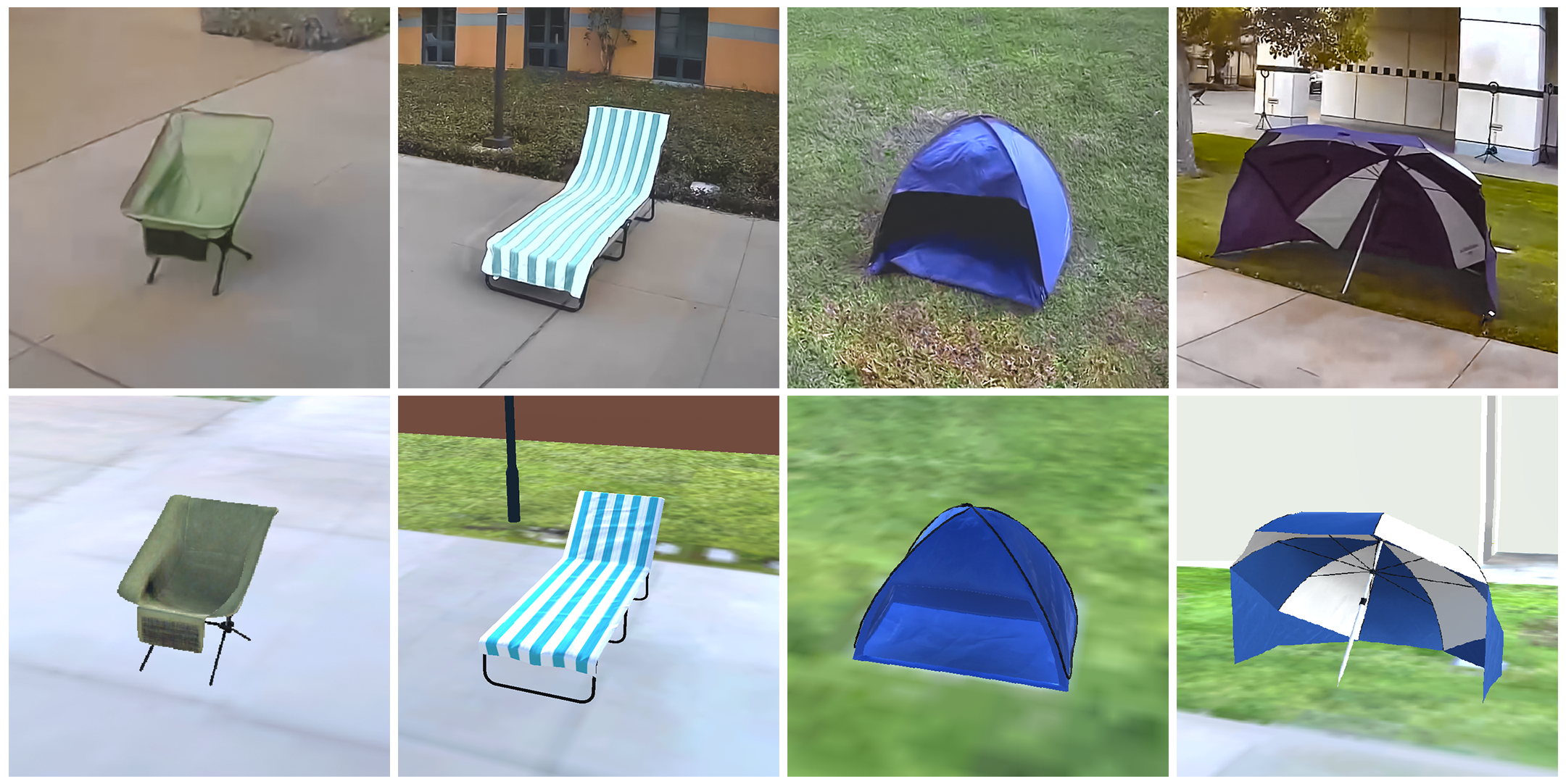}
 \caption{Physical objects (top row) and their virtual replica or twin object. (bottom row). }
 \label{fig:twins}
\end{figure}

    
    \paragraph{Gem location} Each gem in the space had one of three possible locations relative to other objects in the scene: occluded by a physical object/structure (referred to as `physically occluded gems'), occluded by a virtual object/structure (referred to as `virtually occluded gems') and not occluded by any objects (`floating gems'). We consider a gem to be occluded by an object if it is within 50cm of the object and obscured by the object from at least one viewing angle. The floating gems were positioned in open spaces and not in close range of or obscured by any objects in the scene. There were a total of 24 gems in the space, evenly distributed across the three possible locations. A virtual replica (``twin") of each of the 4 types of physical objects was also present in the virtual scene (shown in Figure \ref{fig:twins}) at a different location than the original. To ensure that the actual physical position of the objects did not affect the gem classification performance, we exchanged the positions of the physical objects and their virtual replicas in the scene for half of the participants.

    \paragraph{Lighting} The experiment was considered to be conducted in the `natural' lighting condition when there was sufficient ambient natural light in the environment, but no direct sunlight. This was the hour and a half before sunset. The `night' lighting condition was conducted one hour after sunset with minimal natural light present. Existing area lights and twelve portable LED lights stands were used to illuminate the area enough to enable safe exploration of the space. All night-time LED light sources were directed at physical infrastructure and distributed across the environment for indirect lighting.

    We used different brightness values for the two conditions, adjusted based on our pilot studies. In natural light the maximum brightness was used, and at night the minimum brightness was used. We measured environment light and HoloLens-2 display brightness for a representative object from the participant’s eye position relative to the headset. Ten measurements each at dusk and nighttime, at the mid-point time of each study, were acquired in two locations at opposite ends of the environment. Measurements were made avoiding direct light sources, looking parallel to the ground. The illuminance measurement for natural ambient light reads (M = 1681.5, SD = 448.4) lux while the projected objects emit (M = 2160.6, SD = 460.6) lux. Nighttime ambient light reads (M = 53.6, SD = 12.6) lux while the projected objects emit (M = 137.2, SD = 54.6) lux in our illuminance measurement. The brightness of projected objects was adjusted to avoid change in the perceived distance of the projected objects relative to physical objects, as brighter objects appear closer than similar-sized dimmer objects~\cite{singh2018effect}. The outdoors nighttime illuminance measurements are in the same range as previous experiments in indoor settings, and allow for a better outdoor experience since they are under 10,000 lux~\cite{erickson2020review, kim2019effects}.


Equal numbers of participants were assigned to the lighting conditions. The order of the auditory task conditions was counterbalanced across participants. All participants were presented with equal numbers of gem locations (i.e., physically occluded / virtually occluded / floating).

\subsection{Apparatus}\label{apparatus}
The experiment was conducted in a large outdoor courtyard with 1456 sq.m. (15,672 sq.ft.) of augmented walkable area, including 185 sq.m. (1991 sq.ft.) of lawn area that the participants were discouraged (but not prohibited) from walking on. Participants were instructed to stay within the augmented perimeter, and all participants adhered to this rule. In addition to existing physical structures and objects in the area (e.g. light poles, columns, trash cans, trees and shrubbery), 14 physical objects (4 camping chairs, 3 beach cots, 4 shade tents and 3 beach umbrellas) were also added to the space for the duration of the experiments. For the experiments conducted in the night lighting condition, twelve temporary ring lights were used to ensure even lighting across the space.

The experiment was conducted using a Microsoft HoloLens-2 Mixed Reality headset. 
Participants used a Bluetooth gamepad controller to classify the gems in the scene and respond to the audio task. To reduce the computational overload, we used low-poly 3D models with levels of detail. We use the Hololens provided single-pass instanced rendering and Vertex-Lit shading.

We created a custom web application using Google's web platform Firebase that allowed the experimenters to monitor the user's progress in the study from a distance and ensure that all trials were completed successfully. This application also recorded position and gaze data and all participant interactions with the Bluetooth controller, for playback and performance analysis. The experimenters controlled the study using a Bluetooth keyboard paired with the headset.

\subsection{Participants}

Participants were 48 adult volunteers (27 males, 43 right handed, 24 with vision correction). The ages of the participants ranged from 18 to 61 years (M=27, SD=8.27). Participants were compensated at a rate of \$15 per hour.  All participants reported having normal or corrected-to-normal vision.

\subsection{Procedure}\label{procedure}

Each participant first signed an informed consent form and completed a pre-questionnaire that collected demographic information, sense of direction (Santa Barbara Sense of Direction Scale; SBSOD\cite{hegarty2002development}), previous experience with the outdoor scene, and experience with AR/VR and gaming. The participant's eye gaze was calibrated with the study headset, following which they went through a ten-minute interactive training session that introduced them to the two tasks they had to complete (gem classification and auditory target detection task) and showed them the space the study would be conducted in.

\begin{figure}[t]
 \centering \includegraphics[height=6.52cm]{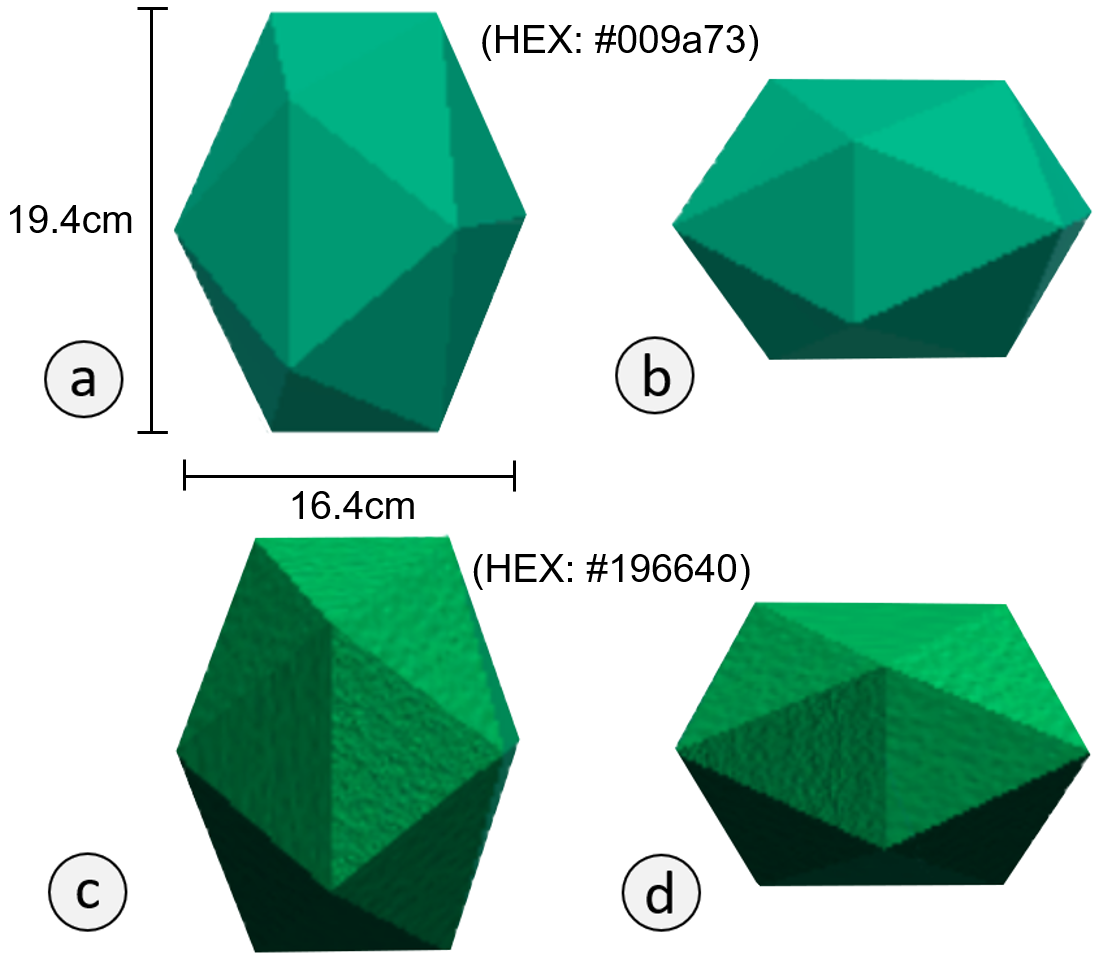}
 \caption{Our four different types of gems: (a) vertical and smooth, (b) horizontal and smooth, (c) vertical and rough, (d) horizontal and rough. Gems were entirely virtual and statically lit . All the gems are green colored, but due to the applied texture the smooth gems appears slightly lighter than the textured ones}
 \label{fig:gems}
\end{figure}

  In the gem classification task, participants walked around the outdoor space and classified each gem they found into one of four categories based on the elongation (vertical or horizontal) and texture (rough or smooth) of the gem (see Figure \ref{fig:gems}). Participant responses were recorded by a button press on the controller, with a different button corresponding to each classification category. As mentioned in the Design section, when the audio task was present, participants pressed a separate button on the control when they detected their randomly assigned target word.

After successfully completing the tutorial, the participant was brought to the study location and provided with a headset and controller. Each participant completed two blocks of three trials each, one block with the audio task present and the other with the audio task absent. The order of the blocks was counterbalanced between participants. A trial was considered complete when the participant decided they had classified all the gems in the space, and thus did not have a fixed duration. The participant was given a 10 second break after each trial, and a 60 second break between the first and second blocks.

After finishing six trials, participants completed a post-study questionnaire where they evaluated their performance in the tasks and provided feedback on the ergonomics of the experience and lighting condition. The entire procedure took approx. 1.5 hours per participant.

 
 \begin{figure}[t]
 \centering
\includegraphics[height=6.52cm]{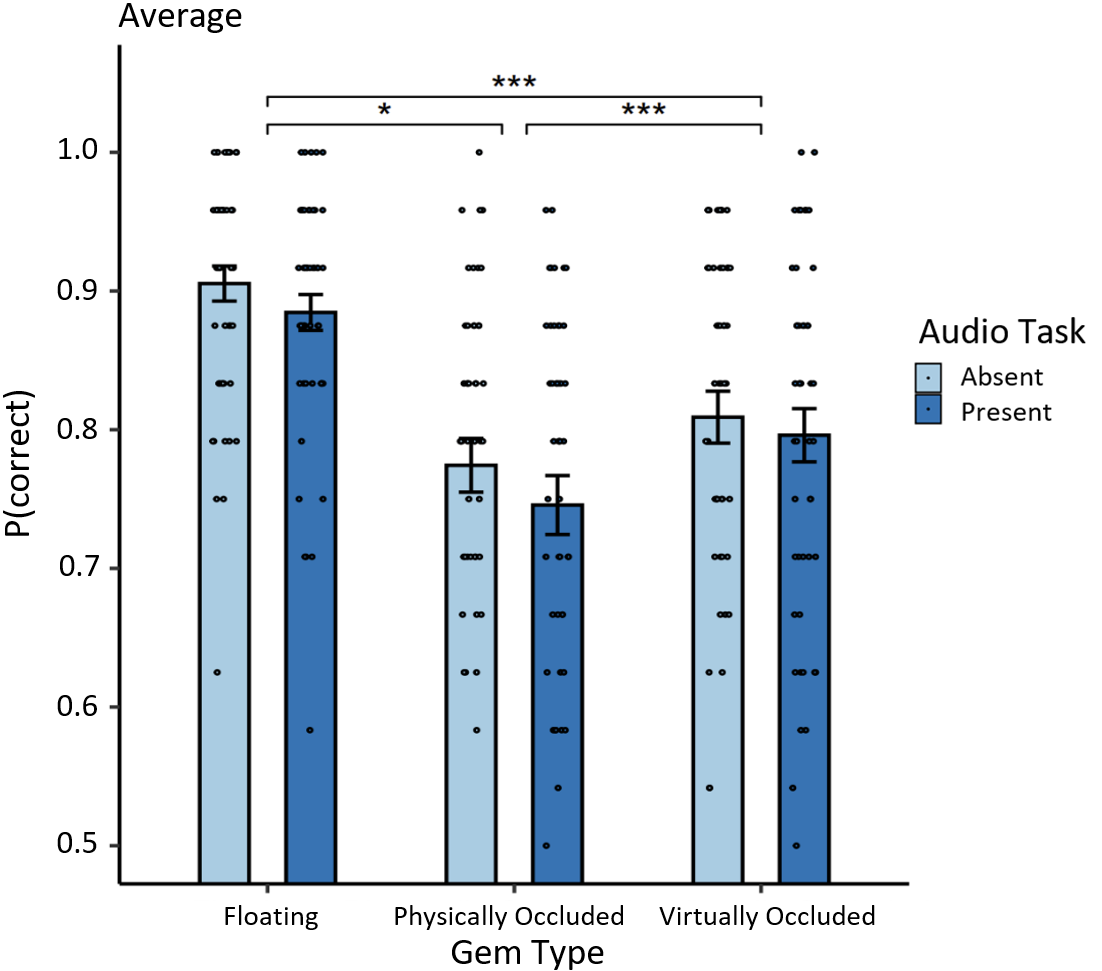}
 \caption{Proportion of correctly discriminated gems (\textit{p(correct)}) plotted as a function of gem location (Physically Occluded, Virtually Occluded, Floating), and audio task (Absent, Present). For this and all following charts: Error bars = SEM. Black dots represent individual data points. Significance bars annotated with:  *: $<$0.05, **:$<$0.01, ***$<$0.001.}
 \label{fig:correctGems}
 \end{figure}

\subsection{Assessment}\label{assessment}

Participants' physical location in the environment, head position, head rotation, eye gaze, and manual responses were continuously recorded for the entire duration of each trial. This information was used to evaluate the participants' performance either directly from these recordings or through the use of a playback software application (see Figure \ref{fig:dataCollection}) that was developed to replay all user interactions and movements by scrubbing back and forth or replaying the recordings at varying speeds. This tool allowed for visual inspection and validation of all recorded data. Some metrics such as total travelled distance and mean head rotation were calculated using the playback software.
 
Task performance was measured in both the gem task and the audio task. In the gem task, participants' performance was measured by the accuracy of classification of gems found in each of the three possible locations (physically occluded, virtually  occluded and floating). This was computed by dividing the number of correct orientation/surface texture discriminations by the number of gems at each of the three possible locations (i.e., 8; p(correct)). A gem was considered to be the target of classification if it was the nearest gem to the participant that was within 4 metres of them and less than 40 degrees of the camera normal. Participants were instructed to look directly at the target during classification, but were given no feedback. The experimenters reviewed all of the participants’ performance in our playback software, and corrected all misclassifications prior to further analysis. Performance in the auditory task was measured as the percentage of target words correctly responded to, where a response was determined to be correct if it occurred after the target word but before the next word in the stream of words. In addition, the time to make correct target detections was measured as the time from the onset of the target word, until the response was made i.e. response time (RT).

\begin{figure}[t]
\includegraphics[width=7.5cm]{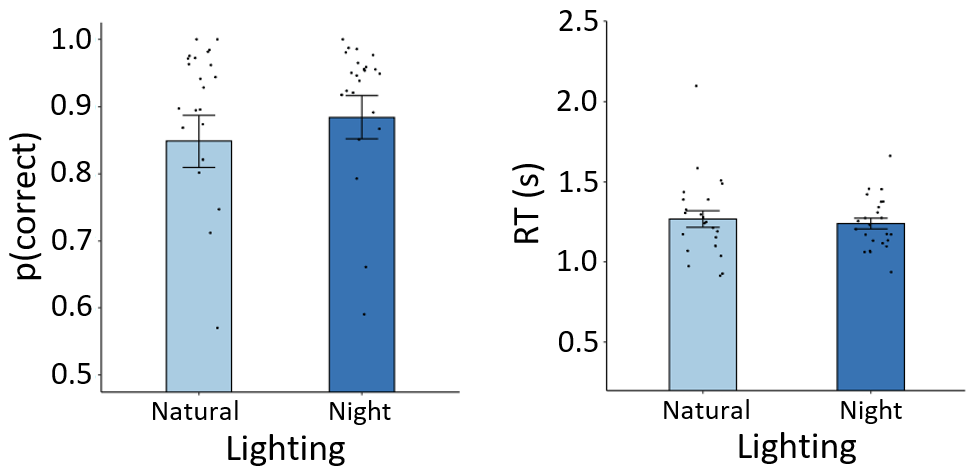}
\caption{Auditory task accuracy (\textit{p(correct)}), represented as proportion of correct responses (left) and response time RT (right) plotted as a function of lighting condition.}
\label{fig:audioPerf}
 \end{figure}

Global behavioral metrics assessed different aspects of the participants' behavior, not explicitly tied to gem and audio target responses when the audio task was absent versus present. These included total head rotation as measured using the unit quaternion distance norm (computed by dividing the  accumulated quaternion distance norm\cite{huynh2009metrics} accumulated over the duration of the task condition by the amount of time taken to complete the task); total distance traveled (meters) during the task condition; and total time elapsed (seconds) when performing the specific task condition.

 To test the impact of our independent variables on our task performance and global behavioral dependent variables, we applied mixed factor ANOVAs using the  {\small rstatix} package in R, where lighting (Natural, Night) was entered as between-subjects and all others variables as within-subjects. ANOVAs were checked for violations of sphericity and homogeneity and the Greenhouse-Geisser correction was applied when necessary. ANOVAs where the assumption of normality was violated were re-analyzed using Friedman's test, a non-parametric equivalent of repeated measures ANOVA. Results remained consistent, so the initial ANOVAs are reported. Furthermore, ANOVA is known to be robust to violations of normality \cite{miller1997beyond}. Effect sizes for the ANOVAs were computed using $\eta_{\mathrm{p}}^{\mathrm{2}}$. Bonferroni corrected pairwise comparisons were used to follow-up significant main effects and interactions. Auditory task performance was compared between natural light and nighttime participants, requiring independent samples t-tests. Effect sizes for the t-tests were quantified using Cohen's $d$. The typical guidelines for interpreting effect sizes are: small ($\eta^{\mathrm{2}}$ = 0.01, $d$=0.2), medium ($\eta^{\mathrm{2}}$ = 0.06, $d$=0.5), and large ($\eta^{\mathrm{2}}$ = 0.14, $d$=0.8) (e.g., \cite{lakensEffects}).

Participants also completed questionnaires prior to, and after, participating in the study. Prior to the study, participants completed a questionnaire that collected demographic information.
Immediately after the study, the participants completed a post-questionnaire designed to assess their observations about the real and virtual environment, obtain feedback on their experience with respect to the ergonomics of the study and lighting conditions, and also record their estimation of their performance in the tasks.

A Mann-Whitney U test was applied to assess the effects of lighting conditions on participants subjective ratings. This non-parametric test was chosen because of the ordinal nature of the Likert rating scales and the data were not normally distributed. The function {\small wilcox.test} from R was used to assess the effect of lighting on 24 dependent variables from the post-questionnaire.

\section{Results}
Our dependent variables included task performance measures, global behavioral metrics and subjective reports from participants. The results from our analyses of these metrics are reported below.

\subsection{Task Performance}

Performance was evaluated in both the gem task and auditory target detection task.

\subsubsection{Gem Task}

Gem task performance is plotted as a function of lighting condition, gem location and auditory task in Figure \ref{fig:correctGems}.  A 2[lighting condition: day, night] x 3[gem location: physically occluded, virtually occluded, free-floating] x 2[auditory task: present, absent] mixed ANOVA computed for gem task accuracy revealed a main effect of gem location (\emph{F}(2,92) = 43.46, \emph{p} $<$.001, $\eta_{\mathrm{p}}^{\mathrm{2}}$ = .49, $large$). None of the other main or interaction effects achieved statistical significance (all \emph{p}-values $>$ .05).   

Pairwise comparisons computed to investigate the main effect of gem location revealed that participants were more accurate at detecting and discriminating free-floating gems when compared to gems hidden behind both physical and virtual objects [\emph{t}(47) = 10.20, 6.30, \emph{p} $<$ .001, \emph{d} = 1.36, 0.95, respectively, $large$].  This result is not surprising, given that free-floating gems were clearly visible from multiple angles and thus less likely to be obscured from view.  Participants were also more accurate at detecting and discriminating gems hidden behind virtual objects when compared to physical objects [\emph{t}(47) = -2.56, \emph{p} = .04,  \emph{d} = 0.36, $small$].  This may be the result of a bias towards virtual objects either because of the relative visibility of those objects or because the target gems were only virtual (see Subjective Ratings result).


\subsubsection{Auditory Target Detection Task}

Auditory task performance is plotted as a function of lightning condition in Figure \ref{fig:audioPerf}.  Overall auditory target detection accuracy was high (overall $p$ correct = .87) and was not modulated by lighting condition ($t$ (46)=-0.70, $p$ =.49, $d$=.20, $small$). Mean response time also did not change as a function of lighting condition ($t$(46)=.46, $p$=0.65, $d$=0.13). These results suggest that the dual-task load was similar under both lighting conditions, although one possible explanation for the lack of effects might be that auditory task accuracy was near ceiling.

\subsection{Global Behavioral Metrics}
The three behavioral metrics analzyed were total head rotation, mean time taken to complete the task(s) and the mean distance traveled.

\subsubsection{Head Rotation} 
Mean unit head rotation during the task is plotted as function of lighting and task condition in Figure \ref{fig:rotation}a. A 2 [lighting condition: day, night] x 2 [auditory task: present, absent] mixed ANOVA revealed a main effect of auditory task on total participant head rotation during the task ($F$(1,46) = 5.82, $p$ = .02,  $\eta_{\mathrm{p}}^{\mathrm{2}}$ = .11, $medium$), such that head rotation was reduced when the auditory task was present [$t$(47) = 2.4, $p$=.02, $d$ = 0.38, $small$].  This result suggests that the dual task condition consumed additional cognitive resources and this reduced participants' scanning of the visual environment. 

There were no significant effects of lighting condition ($F$(1,46) = .84, $p$=.37,  $\eta_{\mathrm{p}}^{\mathrm{2}}$ = .018, $small$) or the interaction ($F$(1,46) = 1.37, $p$ = .25,  $\eta_{\mathrm{p}}^{\mathrm{2}}$ = .029, $small$). 


\subsubsection{Distance Travelled}
Mean distance travelled in meters is plotted as a function of lighting and task conditions in Figure \ref{fig:rotation}b. The results of the mixed model ANOVA did not reveal a significant interaction between audio task and lighting ($F$(1, 46)=.30, $p$=0.59, $\eta_{\mathrm{p}}^{\mathrm{2}}$=0.006). There were also no main effects of audio task or lighting condition (task: ($F$ (1, 46)=3.34, $p$ =0.07, $\eta_{\mathrm{p}}^{\mathrm{2}}$=0.068, $medium$; lighting: ($F$(1, 46) = 2.60, $p$ =0.11, $\eta_{\mathrm{p}}^{\mathrm{2}}$=0.054, $small$)).  One explanation for the lack of effects might simply be that participants were given unlimited time in which to complete their search, which likely contributed to the high variability in distance travelled across subjects, possibly masking any effects of our experimental manipulations.


\subsubsection{Total Time}
Mean total time taken to do the task is plotted as function of lighting and task conditions in Figure \ref{fig:rotation}c.  Consistent with the results for distance travelled, a mixed ANOVA computed for the total time data revealed no main or interaction effects (all $p>$ .05).  


\begin{figure}[t]
\centering \includegraphics[width=6.8cm, keepaspectratio]{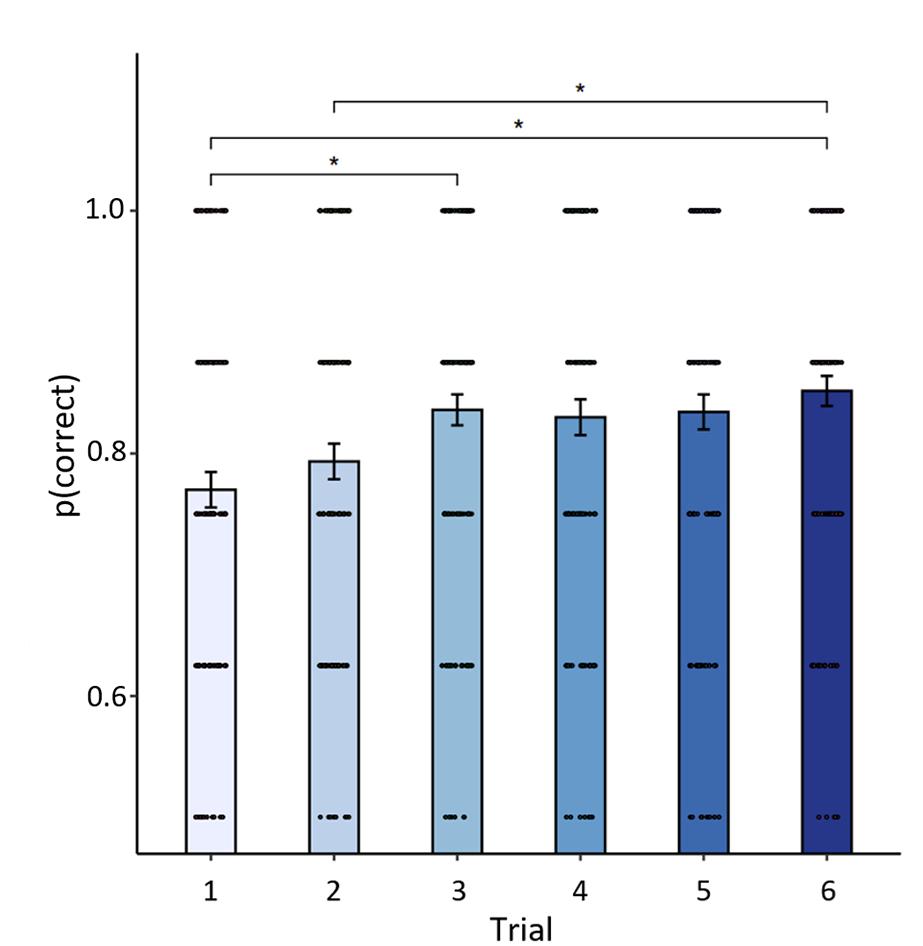}
 \caption{Proportion of correctly discriminated gems (\textit{p(correct)}) plotted as a function of trial order (1-6).}
 \label{fig:learning_acc}
 \end{figure}

\subsection{Learning Effects}
Learning effects on all the task performance and global behavioral metrics were also analyzed. Each participant completed 6 trials, numbered in the sequence they were completed (trial 1 was completed first, and trial 6 completed last).
To assess learning effects on the gem task a 2 [lighting condition: day, night] x 3 [gem location: physically occluded, virtually occluded, floating] x 6 [trial: 1, 2, 3, 4, 5, 6] mixed ANOVA was computed for gem task accuracy. There was a main effect of trial number (\emph{F}(3.83,175.99) = 6.81, \emph{p} $<$ .001, $\eta_{\mathrm{p}}^{\mathrm{2}}$= 0.13, $medium$; see Figure \ref{fig:learning_acc}). Pairwise comparisons revealed that gem task accuracy improved over time (e.g., trial 1 vs. trial 6: \emph{t}(47) = -4.75, \emph{p} $<$ .001, $d$ = 0.78, $medium$). Improved performance over the course of the session likely reflects a combination of participants learning the environment and becoming familiar with repeating gem locations as well as them becoming increasingly comfortable operating in the AR environment. The main effect of gem type was statistically significant as previously reported (\emph{F}(2,92) = 43.46, \emph{p} $<$.001, $\eta_{\mathrm{p}}^{\mathrm{2}}$ = .49, $large$) but no other main effects or interactions were significant (\emph{p} $>$ .05).

\begin{figure}[t]
\centering \includegraphics[width=6.8cm, keepaspectratio]{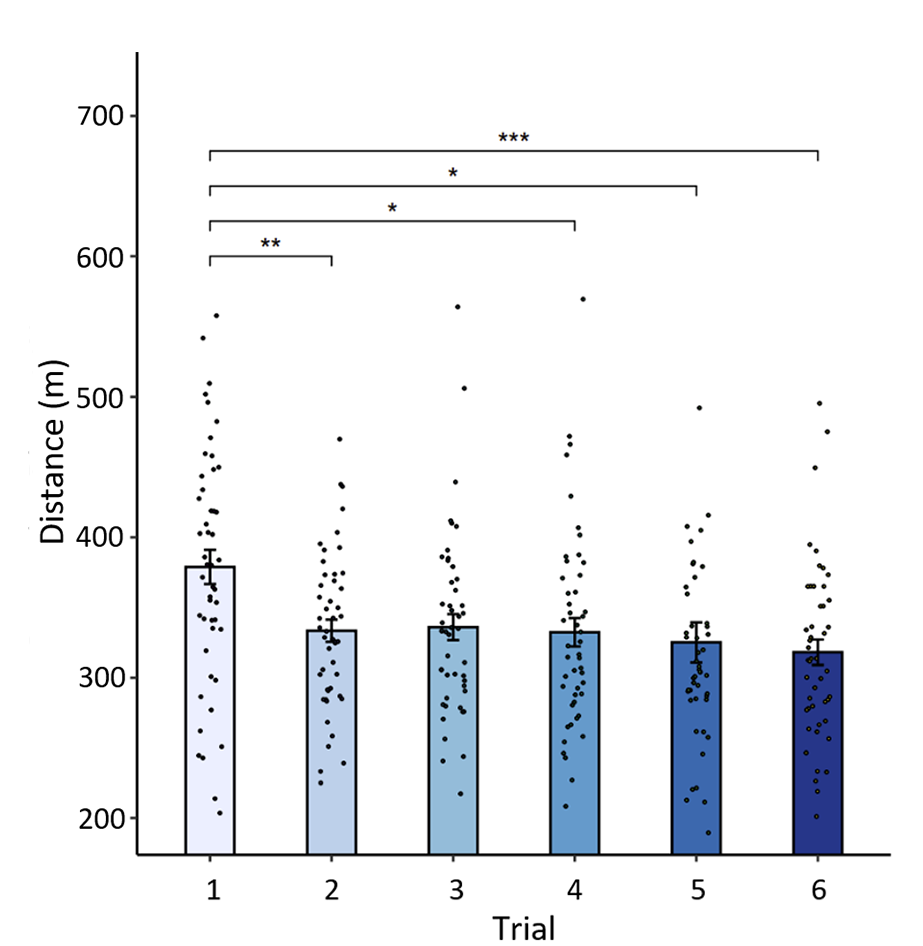}
\caption{Mean distance traveled plotted as a function of trial order.}
\label{fig:learning_distance}
\end{figure}

\begin{figure}[t]
\centering \includegraphics[height=7.2cm, keepaspectratio]{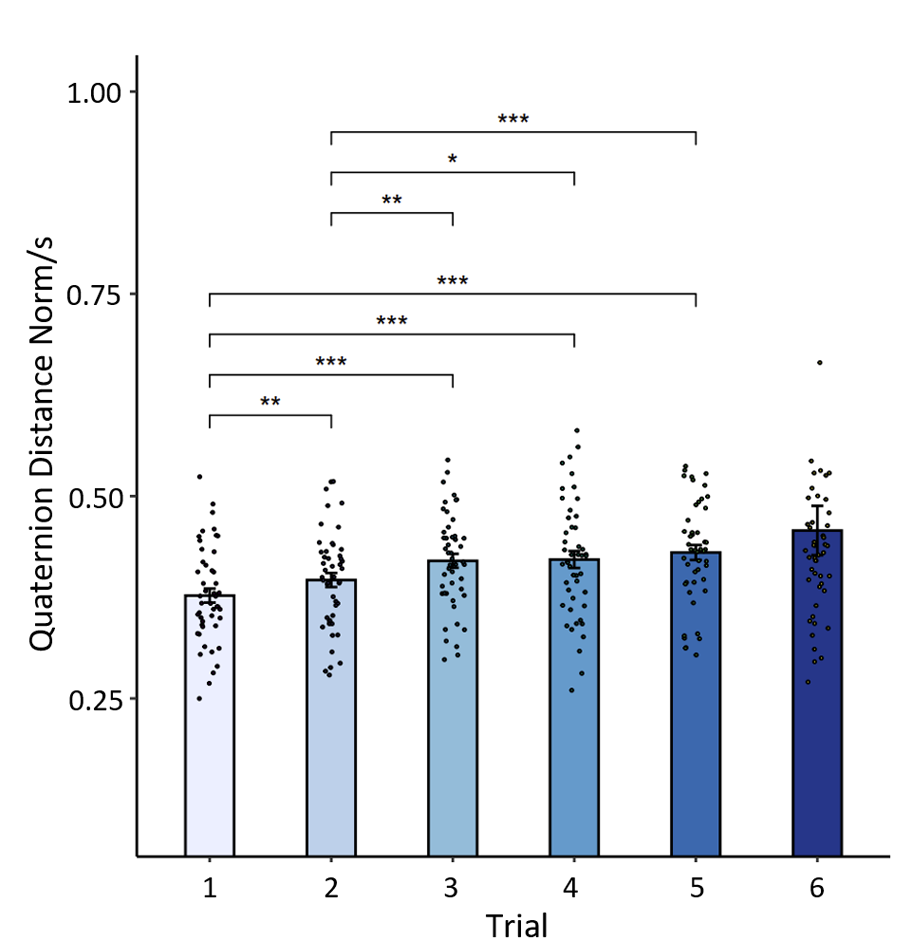}
\caption{Mean unit rotation plotted as a function of trial order.}
\label{fig:learning_rotation}
\end{figure}

\begin{figure}[t]
\centering \includegraphics[height=7.2cm, keepaspectratio]{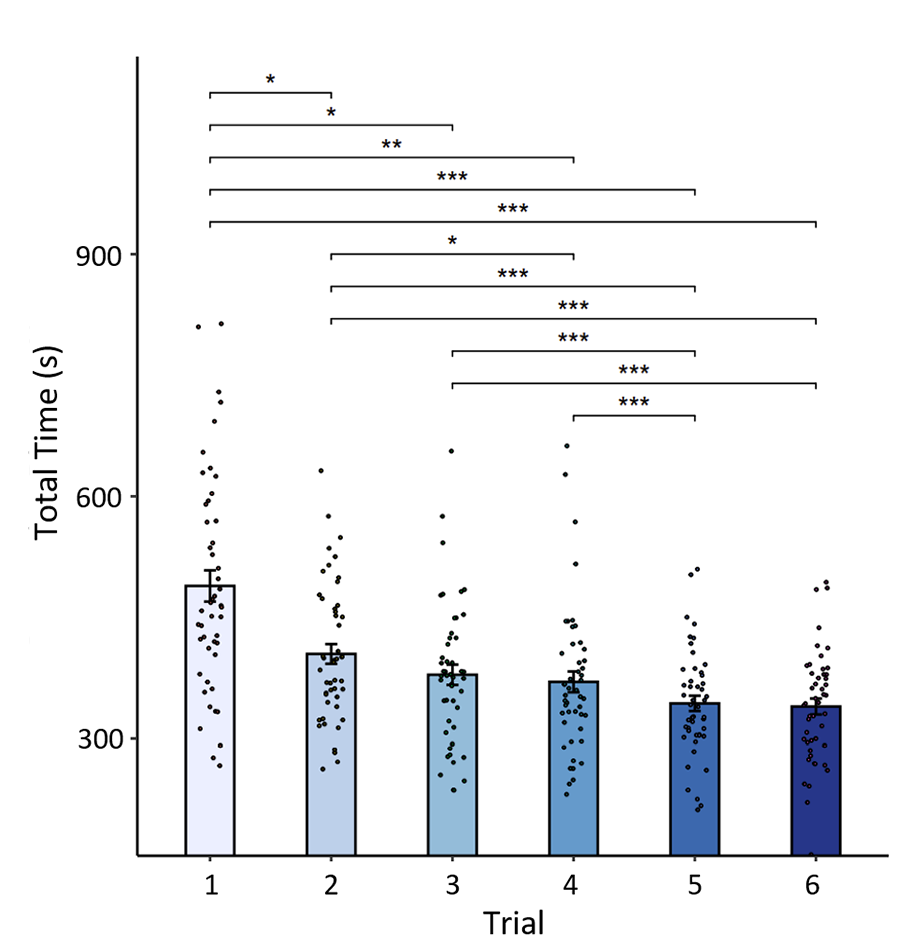}
\caption{Mean elapsed time plotted as a function of trial order.}
\label{fig:learning_time}
\end{figure}

Separate 2 [lighting condition: day, night] x 6 [trial: 1, 2, 3, 4, 5, 6] mixed ANOVAs were conducted to assess learning effects on three dependent variables: distance travelled, total head rotation, and total session time. First, the ANOVA examining distance travelled revealed a main effect of trial number ($F$(3.12,143.47) = 7.20, $p$ $<$ .001, $\eta_{\mathrm{p}}^{\mathrm{2}}$= 0.14, $large$, see Figure \ref{fig:learning_distance}). Pairwise comparisons revealed that the distance participants travelled decreased over the course of the experiment e.g., trial 1 vs. trial 6: ($t$(47) = 5.33, $p$ $<$ .001, $d$ = .82, $large$). The main effect of lighting condition and the interaction were not significant. Second, the ANOVA examining unit head rotation throughout the experiment revealed that there was a main effect of trial number ( $F$(1.23, 56.76) = 4.90, $p$ = .024, $\eta_{\mathrm{p}}^{\mathrm{2}}$= 0.10, $medium$; Figure \ref{fig:learning_rotation}). Pairwise comparisons revealed that unit head rotation increased over the course of the experiment e.g., trial 1 vs. trial 5: ($t$(47) = -6.68, $p$ $<$ .001, $d$ = -0.52, $medium$). The main effect of light and the interaction were not significant. Third, the ANOVA examining elapsed time revelaed a main effect of trial number ($F$(2.70,124.08) = 32.66, $p$ $<$ .001, $\eta_{\mathrm{p}}^{\mathrm{2}}$= 0.42, $large$; Figure \ref{fig:learning_time}). Pairwise comparisons revealed that the time taken to complete each trial decreased over the course of the testing session e.g., trial 1 vs. trial 6: ($t$(47) = 7.74, $p$ $<$ .001, $d$ = 1.39, $large$). The main effect of light and the interaction were not significant. When considered together, these learning effects indicate that participants became increasingly efficient at the gem classification task over the course of the session as they learned the environment and became familiar with likely gem locations.

\subsection{Subjective Ratings and Participant Feedback}

A Mann-Whitney U test was computed to assess the impact of lighting conditions on the post-study questionnaire responses (24 items).  The test yielded three significant effects of lighting condition at the $p$ $<$ 0.05 level in response to the questions: "How comfortable was it to wear the AR headset?", “Please rate the visibility of physical (real) objects?” and, “Please rate the visibility of virtual objects (seen only through the AR headset)?”. Participants reported that the AR headset was significantly more comfortable to wear at night ($Mdn$ = 6) compared to natural light ($Mdn$ = 5), ($U$ = 178.00, $Z$ = -2.33, $p$ = 0.02, $\eta_{}^{\mathrm{2}}$ =.78, $large$). Participants felt that real objects were significantly more visible in natural light ($Mdn$ = 7) than the night condition ($Mdn$ = 6), ($U$ = 395.00, $Z$ = -2.34, $p$ = 0.019, $\eta_{}^{\mathrm{2}}$ =.79, $large$), while they also reported that virtual objects were significantly more visible at night ($Mdn$ = 7) than during the natural light condition ($Mdn$ = 6), ($U$ = 186.50, $Z$ = -2.21, $p$ = 0.027, $\eta_{}^{\mathrm{2}}$ =.70, $large$). 

Our tests yielded significant effects of lighting condition on two of the post-study questions at the $p$ $<$ 0.01 level: "How likely were you to walk into a physical object?" and “Do you think you found all the gems in each scene?”. Participants reported that they were more likely to walk into physical objects in the night ($Mdn$ = 2) than in natural light ($Mdn$ = 1), ($U$ = 142.00, $Z$ = -3.15, $p$ $<$ 0.01, $\eta_{}^{\mathrm{2}}$ = 1.44, $large$). Participants also felt that found more gems in the scene in natural light ($Mdn$ = 6) compared to the night condition ($Mdn$ = 5.5), ($U$ = 404.00 $Z$ = -3.15, $p$ $<$ 0.01). Users actually found all gems in only 22 out of the (48x6=) 288 trials that were conducted: 8 times at night and 14 times during the day.

There was one marginally significant effect of lighting condition for the question "How much fatigue did you experience after the experience?" such that participants reported more fatigue when participating in the night ($Mdn$ = 5.50) versus the natural light condition ($Mdn$ = 4.50), ($U$ = 207.00, $Z$ = -1.69, $p$ = 0.091, $\eta_{}^{\mathrm{2}}$ =.41, $large$).

Our results indicate that participants had increased comfort, decreased fatigue and better visibility of virtual objects at night. However, participants reported that they found more of the gems in natural light than in the night condition. These results demonstrate that ambient lighting conditions can have considerable implications for AR user acceptance and the cognitive-perceptual experience. These results suggest that context switching between virtual and physical objects is associated with increased fatigue in an environment~\cite{gabbard2019effects, arefin2022effect} where both objects are visible, such as our natural light condition. 



\subsubsection{Recall of Objects in the Environment}

Participants were also tested on their recall of the environment by classifying each of a list of objects into one of four categories: absent, present as a real object, present as a virtual object, and present as both a real and virtual object.
Bonferroni corrected pairwise comparisons revealed that participants accurately recalled virtual objects ($t$(47) = -7.03, $p<$0.001, $d$ = -1.37, $large$) and objects that had both virtual and real counterparts ($t$(47) = -4.81, $p<$ 0.001, $d$ = -0.95, $large$) significantly more often than objects that were only present as real objects. They also recalled virtual objects significantly more($t$(47) = -3.21, $p$=0.014, $d$ = -0.65, $medium$) than they correctly recognised the absence of objects. Interestingly, recall did not differ between the two lighting conditions ($F$(46) = 0.071, $p$ = 0.79, $\eta_{\mathrm{p}}^{\mathrm{2}}$=0.002), which suggests that participants were less aware of their physical environment due to immersion in the virtual scene.

\section{Discussion}
The purpose of this study was to investigate user behavior when engaged in a search task in a large outdoor augmented environment. By manipulating the location of target objects, the lighting conditions when engaged in the task, and whether the search task was conducted in isolation or simultaneously with an audio task, we were able to assess the perceptual and cognitive demands affecting performance. Detailed pre-study and post-study questionnaires also assessed subjective ratings of the experience. Together these measurements indicated that wide-area AR tasks are more demanding in natural light compared to at night, gem placement in the environment plays a key role in determining performance, and the cognitive demands imposed by a dual task influenced search patterns. The implications of these findings will be discussed in turn.

\paragraph{Implication 1} The collection of findings seems to indicate that wide-area AR tasks are more demanding in natural light than at night. This conclusion is supported by the subjective ratings that indicated significantly less comfort and a trend towards more fatigue when engaged in the AR task in natural light. Given that the luminance values for our lamp-lit night-time condition are roughly comparable to lamp-lit indoor environments, and also given that the overwhelming majority of AR user studies has thus far been conducted indoors, this highlights the need for additional outdoor AR studies in natural light in order to correctly assess perceptual and cognitive impact.   

\begin{figure}[t]
\centering \includegraphics[width=6.8cm, keepaspectratio]{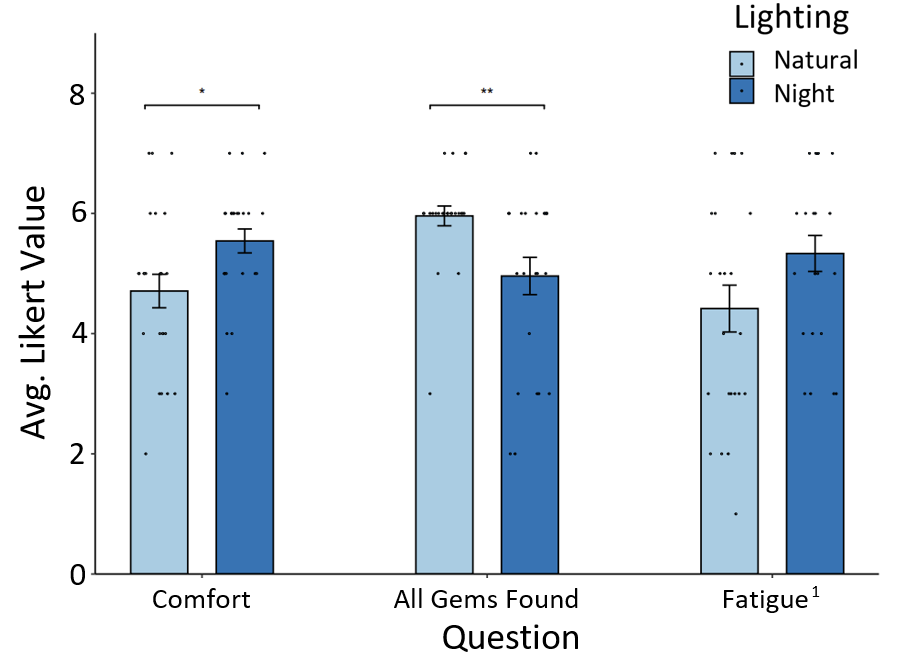}
\caption{Subjective ratings on post-questionnaire items related to Comfort, Fatigue, and perceived number of gems found. 1. Fatigue
item scale was: 1=’A lot of fatigue’ to 7=’No fatigue at all’.}
\label{fig:comfort}
\end{figure}

\begin{figure}[t]
\centering \includegraphics[height=5.3cm, keepaspectratio]{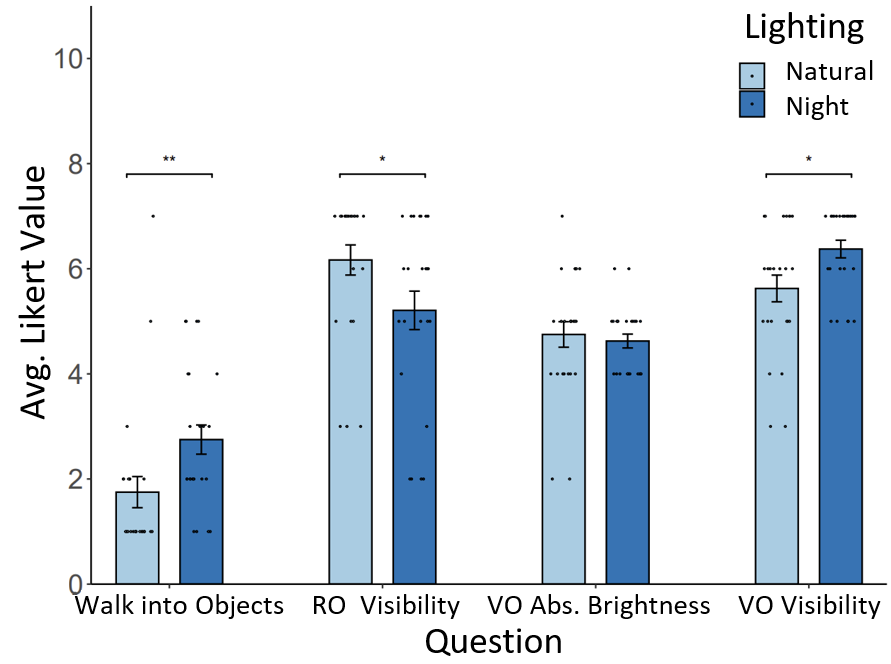}
\caption{Subjective ratings on post-questionnaire items related to virtual object brightness, visibility and likelihood of walking into a physical
object. VO: Virtual Object; RO: Real Object; Abs.: Absolute.}
\label{fig:brightness}
\end{figure}


The subjective reports of virtual object visibility indicate that virtual objects were perhaps more perceptually challenging to discern in the natural lighting condition. The challenges could include dueling focus effects, which may be reduced at night when the virtual objects appear more solid and salient in comparison to the real-world backdrop. The fact that people rated themselves as being more likely to walk into a physical object at night implies that fewer total objects (physical + virtual) were easily visible compared to natural light. The increase in easily visible objects in natural light relative to the night could increase the cognitive load (i.e., more things to attend to) and/or increase the need to segment the virtual and physical worlds. 

\paragraph{Implication 2} Participants' recall of physical objects in the environment was impaired relative to their recall of virtual objects. Reduced object recall could be caused by impaired awareness of, or attention to, the physical environment or by decayed memory or by a combination of both. In the present task, participants were exposed to the environment six times and tested soon after the experiment, thus reducing the likelihood that memory decay is a major contributing factor. Instead, this finding suggests that participants were less aware of the physical environment, perhaps because the target objects were only virtual, or because the novelty effect of AR focused users' attention on the virtual components of the task in general; or because of the differential visibility of the physical and virtual objects (see Implication 1).  

 \paragraph{Implication 3} Task performance was affected by where the gems were located. Floating gems were more easily found and discriminated than those hidden by virtual and physical objects. Methodologically, this is an important validation that the implementation of the occlusion layers was effective. 

 \paragraph{Implication 4} Dual tasking modified behavior. The most robust effect consistent with this conclusion is that participants exhibited less head rotation when doing both the gem task and the audio task compared to the gem task alone. Active search of the environment requires head and eye movements and a reduction of head rotation (i.e., movement) implies a more restricted search strategy. Interestingly, this effect was numerically larger in natural light, which is also when participants trended towards reporting to be more fatigued. 


 
\subsection{Limitations}
The present study had several limitations.  First, a number of the statistical tests did not reveal differences in the expected direction. For example, while dual tasking did affect search strategy, it did not reliably affect performance. The lack of statistically reliable effects was likely driven by the relative ease of the task (i.e., a ceiling effect) and/or insufficient low statistical power resulting from our relatively small sample \cite{richard2022within}. Consistent with the power interpretation, there were a number of instances where analyses revealed medium or greater effect sizes but failed to reach conventional significance thresholds.  Second, tracking inaccuracies occurred during some experimental sessions, leading to errors in the registration of physical occluders and slight discrepancies between the virtual environment that each participant encountered. Participants, however, reported that they found the environment to be generally stable, and experimenters aligned the environment for every user to ensure that registration was as accurate as possible. Third, there were technical limitations associated with the AR headset, such as a limited ability to cast light (and therefore shadows) which affected the rendering of textures. While we were unable to use the headset in bright sunshine due to poor contrast ratios that would affect the users' ability to distinguish the virtual environment\cite{erickson2020review}, the two lighting conditions tested did not have the same problem. Fourth, testing took place outdoors on the UCSB campus, so there was some between-session variability. UCSB is situated in a pocket of Geography known for its stable and reliable weather conditions, but some minor variability remained, as well as some variable interference from pedestrian foot traffic.  Every effort, however, was taken to maximize experimental control by directing pedestrians away from the subject as well as rescheduling experiments when the weather conditions differed from the norm.  Fifth, our study design required that we test participants either in the evening or at night, so it is possible that fatigue or tiredness may have selectively impacted performance, although we note that participant reports actually trended towards feeling less fatigued at night, which is one of the interesting observations to be followed up with additional experimentation.

\begin{figure}[t]
\centering \includegraphics[height=5.3cm, keepaspectratio]{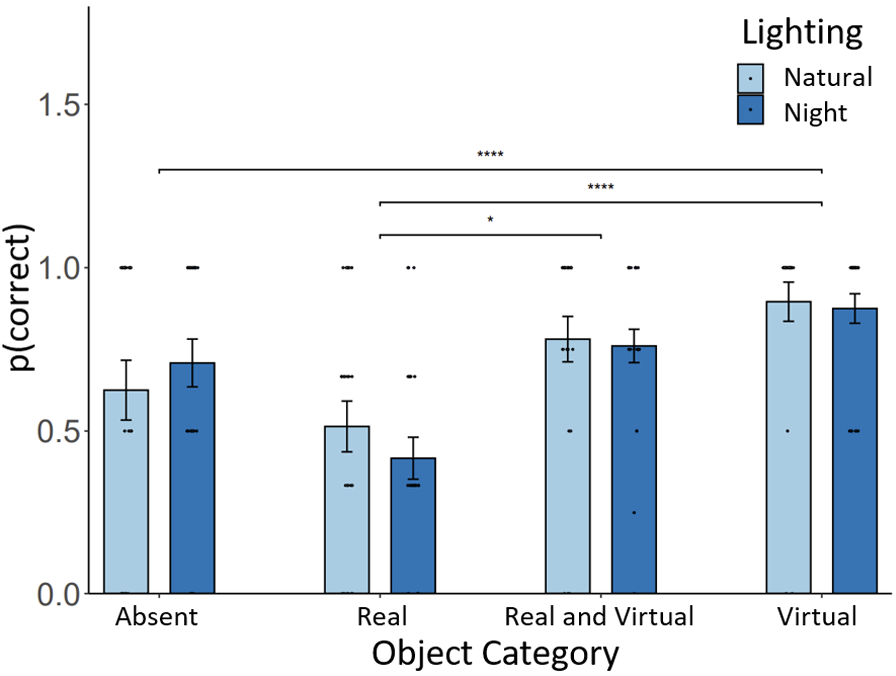}
\caption{Mean accuracy of responses to object recall questions.}
\label{fig:time}
\end{figure}

\section{Conclusion}
Mobile AR has clear applications to many tasks that involve navigation, inspection, and decision-making in a wide variety of physical environments. Our results demonstrate that there are technical, perceptual, and cognitive factors that must be considered if the full potential of anywhere and anytime mobile AR is to be realized. Specifically, this paper presented a wide-area outdoor user study with commercial AR headsets, examining the impact of environmental lighting conditions on AR performance and experience in a controlled experiment ($n$=48). The results regarding increased user comfort and perceptual ease in lower-luminance lamp-lit conditions compared to outdoor natural light raises important questions and can help inform the design of next-generation AR eyewear. Follow-up studies are needed to understand the exact extent and nature of the involved perceptual and psychophysical variables. In particular, the controlled inclusion of {\em physical search targets} could provide insights into the exact nature of the user's relative attention focus on the virtual portion of the AR world. We also would like to further explore the underlying reasons for higher user discomfort during bright daylight, looking, e.g., at the dueling nature of visual focus on augmented physical objects, with users having to switch attention back and forth among physical objects and potentially slightly transparent virtual augmentations.

\section{Acknowledgements}
This work was funded by the U.S. Army Combat Capabilities Development Command Soldier Center Measuring and Advancing Soldier Tactical Readiness and Effectiveness (MASTR-E) program through award W911NF-19-F-0018 under contract W911NF-19-D-0001 for the Institute for Collaborative Biotechnologies. Additional funding came from ONR awards N00014-19-1-2553 and N00174-19-1-0024, as well as NSF award IIS-1911230. The authors thank Natalie Juo and Anabel Salimian for their assistance with data collection.

\bibliographystyle{abbrv-doi}

\bibliography{bibliography}
\end{document}